\documentstyle[12pt]{article}

\begin{document}
\title{$K$-causality and domain theory}
\author{ {\small N. Ebrahimi } \\ {\small Department of Mathematics, Shahid
Bahonar University of Kerman}\\
{\small Kerman, Iran} \\
{\small neda$\_$eb55@yahoo.com}}
\date{}

\maketitle

\begin{abstract} Using the relation $K^{+}$, we prove that a certain type of stably causal spacetimes is a jointly
bicontinuous poset whose interval topology is the manifold
topology.
\end{abstract}

Alexandrov topology.

\vspace{1cm}

\section{Introduction}  It is shown by Martin and Panangaden \cite{MP}
that it is possible to reconstruct globally hyperbolic spacetimes
in a purely order theoretic manner using the causal relation
$J^{+}$. These spacetimes belong to a category that is equivalent
to a special category of domains called interval domains
\cite{ID}. In this paper we use the causal relation $K^{+}$
instead of $J^{+}$. The relation $K^{+}\subseteq M \times M $ is defined as the smallest transitive
closed relation which contains $I^{+}$ \cite{SW}. This definition
arose from the fact that the causal relation, $J^{+}$, is
transitive but not necessarily closed and $\overline{J^{+}}$ is
closed but not necessarily transitive. The spacetime $(M,g)$ is
$K-$causal if $K^{+}$ is antisymmetric. Recently it is proved by Minguzzi that stable
causality and $K$- causality are coincide. In globally hyperbolic and causally simple spacetimes $K^{+}=J^{+}$.
In this paper we prove that $K$- causal spacetimes, in which
$int(K^{\pm}(.))$ are inner continuous are jointly bicontinuous posets.

\section{Preliminaries} A poset is a partially ordered set, i.e,
a set together with a reflexive, antisymmetric and transitive
relation.

 \noindent\ In a poset $(P,\sqsubseteq )$, a nonempty subset
 $S\subseteq P$ is called directed (filtered) if $(\forall x,~y\in
 S)(\exists z\in S)$ $x,~y\sqsubseteq z$ ($(\forall x,~y\in
 S)(\exists z\in S)$ $z\sqsubseteq x,~y$). The supremum(infimum) of $S$
 is the least of its upper bounds (greatest of all its lower bounds)
 provided it exists.

 \noindent\ For a subset $X$ of a poset $P$, set:
 $$\uparrow X= \{ y\in P : (\exists x\in X)~ x\sqsubseteq y
 \},~~~\downarrow X= \{ y\in P : (\exists x\in X)~ y\sqsubseteq x
 \}.$$
 A dcpo is a poset in which every directed subset has a supremum.
  The least element in a poset,
 when it exists, is the unique element $\bot$ with $\bot
 \sqsubseteq x$ for all $x$.

 \noindent\ A subset $U$ of a poset is scott open if:

 \noindent\ (i) $U$ is an upper set: $x\in U$ and $x\sqsubseteq
 y\Rightarrow y\in U$.

 \noindent\ (ii) For every directed $S\subseteq P$  with supremum that
 $\bigsqcup S\in U$ implies $S\cap U\neq \emptyset $.

 \noindent\ The collection of scott open sets on $P$ is called the
 scott topology.

 \noindent\ {\bf Definition 2.1.} For elements $x$, $y$ of a poset,
 write $x\ll y$ if and only if for all directed sets $S$ with a
 supremum,
 $$y\sqsubseteq \bigsqcup S \Rightarrow (\exists s\in S)~ x\sqsubseteq
 s.$$

 \noindent\ We set $\Downarrow x= \{ a\in P : a\ll x \}$ and
 $\Uparrow x= \{ a\in P : x\ll a \}$.

 \noindent\ For symbol "$\ll $", read "way below".

 \noindent\ {\bf Definition 2.2.} A basis for a poset $P$ is a subset
 $B$ such that $B \cap \downarrow x$ contains a directed set with
 supremum $x$ for all $x\in P$. A poset is continuous if it has a
 basis. A poset is $\omega$- continuous if it has a countable
 basis.

\noindent\  {\bf Definition 2.3.} For elements $x$, $y$ of a
poset,
 write $x\ll_{d} y$ if and only if for all filtered sets $S$ with
 an infimum,
 $$\bigwedge S\sqsubseteq x \Rightarrow (\exists s\in S)~ s\sqsubseteq
 x.$$
\noindent\ We set $\Downarrow_{d} x= \{ a\in P : a\ll_{d} x \}$
and $\Uparrow x_{d}= \{ a\in P : x\ll_{d} a \}$.
 \noindent\ For symbol "$\ll_{d} $", read " way above".

 \noindent\ {\bf Definition 2.4.} A poset $P$ is dual continuous if
 $ \Uparrow_{d} x$ is filtered with infimum $x$ for all $x\in P$.

 A poset $P$ is bicontinuous if it is both continuous and dual
 continuous. In addition a poset is called jointly bicontinuous if
 it is bicontinuous and the way below relation coincides with the
 way above relation. A bicontinuous poset is called globally
 hyperbolic poset if all of its intervals, $\ [ a,b\ ]= \uparrow a \cap\downarrow b $, are compact in the interval
 topology.

 \noindent\ {\bf Proposition 2.5.}\cite{D} If $x\ll y$ in a continuous
 poset $P$, then there is $z\in P$ with $x\ll z\ll y$.

 \noindent\ {\bf Definition 2.6.} On a bicontinuous poset $P$, sets of the form
 $$(a,b) :=\{ x\in P : a\ll x\ll_{d} b \}$$

 \noindent\ form a basis for a topology called the interval
 topology.

A useful example of continuous domains is upper space.

\noindent\ {\bf Example 2.7.} Let $X$ be a locally compact
Hausdorff
 space. Its upper space $UX=\{ K\neq \emptyset : K$ is compact $\}$,
 with $A\sqsubseteq B \Leftrightarrow B\subseteq A $ is a
 continuous dcpo. For $K$, $L\in UX$, $K\ll L$ if and only if
 $L\subseteq int(K)$.

\section{ Causal structure of a spacetime}

 \noindent\ In this section we suppose that $(M,g)$ is a spacetime and $I^{+}$
  and $J^{+}$ are the chronological and causal relations
  \cite{GL}. The spacetime $M$ is
globally hyperbolic if it is causal and $J^{+}(x)\cap J^{-}(y)$
is compact for every $x$, $y\in M$. Martin and Panangaden defined
an order on the spacetime $M$ in the following manner:

$$p\sqsubseteq q\equiv q\in J^{+}(p).$$

\noindent\ They proved the following theorem about Globally
hyperbolic spacetimes:

\noindent\ {\bf Theorem 3.1.} If $M$ is a globally hyperbolic
spacetime, then $(M,\sqsubseteq)$ is a biconinuous poset with
$I^{+}=\ll$ whose interval topology is the manifold topology.

\noindent\ This theorem suggests a formulation of causality
independently of geometry. In this paper we try to generalize
theorem 3.1. We use the relation $K^{+}$ instead of $J^{+}$.

\noindent\ If $U$ is an open neighborhood of $M$, then we denote
by $J_{U}^{+}$ the causal relation on the spacetime $U$ with the
induced metric. we recall that every event $p$ of a spacetime
$(M,g)$ admits arbitrary small globally hyperbolic neighborhoods.

 An open set $U$ is $K$- convex if for all $p$, $q\in U$,
$K^{+}(p)\cap K^{-}(q)\subseteq U$ \cite{SW}.

\noindent\ The spacetime $M$ is strongly $K$- causal at $p$ if it
contains arbitrary small $K$- convex neighborhoods of $p$, and it
is strongly $K$- causal if it is strongly $K$- causal for all
$p\in M$. $K$- causality implies strong $K$- causality \cite{SW}.
The converse is trivial.

With $int(B)$ and $\overline{B}$, we denote the topological
interior and closure of $B\subseteq M$, respectively. Let $F$ be a
function which assigns to each point $p\in M$ an open set
$F(p)\subseteq M$. We say that $F$ is inner continuous if for any
$p$ and any compact set $C\subseteq F(p)$, there exists a
neighborhood $U$ of $p$ with $C\subseteq F(q)$, for every $q\in
U$. In globally hyperbolic spacetimes $K^{+}=J^{+}$ and
$int(K^{\pm}(.))=I^{\pm}(.)$ that are inner continuous.

\noindent\ {\bf Lemma 3.2.}\cite{K} In a $K$- causal spacetime
$(M,g)$, $int(K^{+}(.))$ and $int(K^{-}(.))$ are inner continuous
if and only if for every $p,~q\in M$, $p\in
int(K^{-}(q))\Leftrightarrow\ q\in int(K^{+}(p))$.

\noindent\ {\bf Lemma 3.3.} $int(K^{+}(.))$ and $int(K^{-}(.))$ are outer continuous.

\noindent\ {\bf Definition 3.3.} A $K$- causal spacetime $(M,g)$ is called $K$- causally continuous if
$int(K^{+}(.))$ and $int(K^{-}(.))$ are inner continuous.

\noindent\ {\bf Definition 3.4.} Let $(M,g)$ be a spacetime.
Alexandrov topology on $M$ is the one which admits as a base,
$$B_{A}=\{I^{+}(p)\cap\ I^{-}(q):~p,~q\in\ M\}.$$

\noindent\ {\bf Theorem 3.4.}\cite{GL} For a spacetime $(M,g)$,
the following properties are equivalent:

\noindent\ (a) $(M,g)$ is strongly causal.

\noindent\ (b) Alexandrov topology is equal to the original
topology on $M$.

Using the relation $K^{+}$, we define the following topology on
$M$.

\noindent\ $K$- Alexandrov topology is the one with the base,

$$B_{K}=\{int(K^{+}(p))\cap int(K^{-}(q))~:~p,~q\in M\}.$$

\noindent\ {\bf Theorem 3.5.} The following are equivalent:
are equivalent:

\noindent\ (a) $(M,g)$ is $K$- causally continuous.

\noindent\ (b) $K$- Alexandrov topology is equal to the original
topology on $M$.

\noindent\ {\bf Proof.} Assume that $(M,g)$ is $K$- causal. $K$-
causality implies strong $K$- causality. Definition of strong
$K$- causality implies that each point has arbitrary small $K$-
convex neighborhoods. If $V$ be an open neighborhood of $p$ in the
manifold topology, then there exists a causally $K$- convex
neighborhood $U$ of $p$, $U\subseteq\ V$, that is contained in a
globally hyperbolic neighborhood. Indeed,
$K_{U}^{+}(p)=J_{U}^{+}(p)$ and $int(K_{U}^{+}(p))=I_{U}^{+}(p)$.
Thus $K$- Alexandrov topology on $(U,g|_{U})$ agrees with
Alexandrov topology.  Using theorem 3.4 and the fact that a
strongly $K$- causal spacetime is strongly causal demonstrate
that the manifold topology on $U$ agrees with $K$- Alexandrov
topology. Hence $K$- Alexandrov topology agrees with the manifold
topology.

 Conversely, suppose that $M$ is not strongly $K$- causal at $p$. There
is a neighborhood $V$ of $p$ that for every neighborhood
$W\subseteq\ V$ of $p$ there exist points $p',~q'\in\ W$ such
that $K^{+}(p')\cap\ K^{-}(q')$ is not a subset of $W$. Thus
there isn't any open set in $K$- Alexandrov topology that is
contained in V. Indeed, if for $c$, $c'\in M$,
$W=int(K^{+}(c))\cap\ int(K^{-}(c'))\subseteq V$ then by
assumption, there is a point $d\in K^{+}(p')\cap\ K^{-}(q')$ that
$d\notin W$. But since $int(K^{\pm}(.))$ are inner continuous,
$d\in W$ that is a contradiction. As a consequence, $K$-
Alexandrov topology is different from the given manifold topology.

\section{Spacetime and domain theory}
\noindent\ Let $M$ be a $K$- causal spacetime. We write the
relation $K^{+}$ as:

$$p\sqsubseteq\ q\equiv\ (p,q)\in K^{+}.$$

\noindent\ {\bf Example 4.2.} Let $M$ be a globally hyperbolic
spacetime. In a globally hyperbolic spacetime, $K^{+}=J^{+}$. Let
$S$ be a directed set with supremum, then $\bigsqcup
S=\bigcap_{s\in S} \ [ s,\bigsqcup S \ ] $. Let $V$ be an
arbitrary small neighborhood of $\bigsqcup S$. Using the
approximation on the upper space of $M$, $\overline{V}\ll
\bigsqcup S= \bigcap_{s\in S} \ [ s,\bigsqcup S \ ] $ where the
intersection is a directed collection of nonempty compact sets by
directedness of $S$ and global hyperbolicity of $M$. Thus for
some $s\in S$, $\ [ s,\bigsqcup S\ ] \subseteq \overline{V}$.

\noindent\ {\bf Lemma 4.3.} Let $p,~q$ and $r\in M$. Then:

\noindent\ i) $p\sqsubseteq\ q$ and $r\in
int(K^{+}(q))\Rightarrow\ r\in int(K^{+}(p))$.

\noindent\ ii) $p \in int(K^{-}(q))$ and $q\sqsubseteq\
r\Rightarrow\ p\in int(K^{-}(r))$.

 \noindent\ {\bf Lemma 4.4.} Let $y_{n}$ be a sequence in $M$ with $y_{n}\sqsubseteq y$ ($y_{n}\sqsubseteq y$) for all
 $n$ and $lim_{n\rightarrow \infty} ~y_{n}=y$; then $\bigsqcup y_{n}=y$ ($\bigwedge y_{n}=y$).

 \noindent\ {\bf Proof.} Let $y_{n}\sqsubseteq x$ for every $n\in N$. Since $K^{+}$ is
 closed and $y_{n}\in K^{-}(x)$, $y=lim_{n\rightarrow \infty}~y_{n}\in K^{-}(x)$. Thus
 $y\sqsubseteq x$ and this proves $y= \bigsqcup y_{n}$. The proof
 for the dual part is similar to this.

 \noindent\ Note that the above lemma is true for every causal
 closed relation.

 \noindent\ {\bf Lemma 4.5.}\cite{MP} For any $x\in M$, $I^{-}(x)$ ($I^{+}(x)$) contains an
 increasing (decreasing) sequence with supremum (infimum) $x$.

\noindent {\bf Lemma 4.6.} Let $S$ be a directed set in $(M,g)$ with supremum $\bigsqcup S$. Then there is an increasing sequence $\{s_{n}\}$ in $S$ such that $lim_{n\rightarrow \infty} s_{n}=\bigsqcup S$.

\noindent {\bf Proof.} Let $A=\{\{s_{n}\}: s_{n}\in S, s_{n}\sqsubseteq s_{n+1}~ \forall n\in N\}$. We define an equivalence relation on $A$ in the following manner:
$$\{s_{n}\}\sim \{s'_{n}\}\Leftrightarrow \exists~m\in N~: s_{n}=s'_{n}~\forall ~n>m.$$
\noindent Now we define a partial order on $A/\sim$.
$$[\{s_{n}\}]\sqsubseteq _{1} [\{s'_{n}\}]\Leftrightarrow \exists m\in N:~ s_{n}\sqsubseteq s'_{n}, \forall ~ n\geq m.$$
\noindent Suppose that $\{a_{m}\}_{m\in N}=\{[\{s_{m,n}\}_{n\in N}]:m\in N\}$ is a chain in $A/\sim$. We show that it has an upper bound. We define the sequence $\{b_{m}\}$ in the following manner:
$$b_{1}=s_{1,n_{1}}~:~s_{1,n}\sqsubseteq s_{2,n}~\forall n>n_{1},$$
$$b_{i}=s_{i,n_{i}}: s_{i,n}\sqsubseteq s_{i+1,n} \forall n>m ~and~ n_{i}=max\{m,n_{1},...,n_{i-1}\}.$$
\noindent It is easy to show that $[\{b_{m}\}]$ is an upper bound of $\{a_{m}\}$. Hence by zorn's lemma $A/\sim$ has a maximum element $c=[\{c_{m}\}]$. Suppose by contradiction that there is a neighborhood $U$ of $\bigsqcup S$ with compact closure such that $S\cap U=\emptyset$. Let $\{c_{m}\}$ be a representation of $[\{c_{m}\}]$. Since $c_{m}\sqsubseteq \bigsqcup S$, there is  $d_{m}\in \partial U$, such that $c_{m}\sqsubseteq d_{m}$ and $d_{m}\sqsubseteq \bigsqcup S$. $\{d_{m}\}$ has an accumulation point like $d$ since $\partial U$ is compact. There is $m\in N$ such that $c_{n}\sqsubseteq c_{n+1}, \forall n>m$ and $K^{+}$ is closed. Hence $c_{i}\sqsubseteq d_{j}$, $\forall i,~j>m$ and consequently $c_{i}\sqsubseteq d$, $\forall i>m$. But $[\{c_{m}\}]$ is a maximal element of $A/\sim$ and this implies that $d$ is an upper bound of $S$ which is a contradiction to the fact that $d\sqsubseteq \bigsqcup S$ and $d\neq \bigsqcup S$.

\noindent\ {\bf Theorem 4.6.} Let $M$ be a $K$- causally continuous spacetime. Then
$$x\ll\ y\Leftrightarrow\ y\in int(K^{+}(x))\Leftrightarrow x\ll_{d} y.$$

\noindent\ {\bf Proof.} Let $y\in int(K^{+}(x))$. If for the
directed set $S$ $y \sqsubseteq\ \bigsqcup S$, then by assumption
and lemma 3.2, $ \bigsqcup S \in int(K^{+}(x))$.By lemma and the fact that
$int(K^{+}(x))$ is open, there exists $s\in S$ such that $s \in
int(K^{+}(x))$. Consequently, $x\ll\ y$.

\noindent\ If $x\ll\ y$, by lemma 4.5 there exists an increasing
sequence $y_{n}$ in $I^{-}(y)$ such that $\bigsqcup\ y_{n}=y$.
Thus $x\sqsubseteq\ y_{n}$, for some $n$. Since $I^{+}$ is an
open relation, $x\in int(K^{-}(y))$. The proof of the other part
is similar to this.

\noindent\ {\bf Theorem 4.7.} If $M$ is a $K$- causally continuous spacetime, then $(M,\sqsubseteq\ )$ is
a jointly bicontinuous poset with $\ll\ =int(K^{-}(.))$ whose
interval topology is equal to the manifold topology.

{\bf Proof.} By lemma 4.6, $\Downarrow x=int(K^{-}(x)$. In
addition, by lemma 4.5, for every $x\in M$ there is an increasing
sequence $x_{n}\subseteq I^{-}(x)\subseteq
int(K^{-}(x))=\Downarrow x$ with $\bigsqcup x_{n}=x$. Hence $M$
is continuous. In a similar way we can prove that it is dually
continuous.  In addition, by theorem 4.6 and 3.4, interval
topology is equal to the manifold topology.

\end{document}